\documentclass[doublecol]{epl2}
\usepackage{color}
\usepackage{graphicx}
\usepackage{dcolumn}
\usepackage{bm}
\usepackage{array}
\usepackage[table]{xcolor}

\newcommand{\PreserveBackslash}[1]{\let\temp=\\#1\let\\=\temp}
\newcolumntype{C}[1]{>{\PreserveBackslash\centering}p{#1}}
\newcolumntype{R}[1]{>{\PreserveBackslash\raggedleft}p{#1}}
\newcolumntype{L}[1]{>{\PreserveBackslash\raggedright}p{#1}}
\begin{document}

\title{Link prediction in complex networks: a local na\"{\i}ve Bayes model}

\author{Zhen Liu\inst{1} \and Qian-Ming Zhang\inst{1} \and Linyuan L\"u\inst{2}\footnote{Corresponding author:\email{linyuan.lue@unifr.ch}} \and Tao Zhou\inst{1}}
\shortauthor{ZHEN LIU \etal} \institute{ \inst{1} Web Sciences
Center, School of Computer Science and Engineering, University of
Electronic Science and Technology of China,
Chengdu 610054, People's Republic of China\\
\inst{2}Department of Physics, University of Fribourg, Chemin du Mus\'{e}e 3,
Fribourg CH-1700, Switzerland}


\abstract {Common-neighbor-based method is simple yet effective to
predict missing links, which assume that two nodes are more likely
to be connected if they have more common neighbors. In such method,
each common neighbor of two nodes contributes equally to the
connection likelihood. In this Letter, we argue that different
common neighbors may play different roles and thus lead to different
contributions, and propose a local na\"{\i}ve Bayes model
accordingly. Extensive experiments were carried out on eight real
networks. Compared with the common-neighbor-based methods, the
present method can provide more accurate predictions. Finally, we
gave a detailed case study on the US air transportation network.}

\maketitle

\section{Introduction}

The problem of link prediction aims at estimating the likelihood of
the existence of a link between two nodes in a given network, based
on the observed links \cite{Lu2011}. Recently, the study of link
prediction has attracted much attention from disparate scientific
communities. In the theoretical aspect, accurate prediction indeed
gives evidence to some underlying mechanisms that drives the network
evolution \cite{Liu}. Moveover, it is very possible to build a fair
evaluation platform for network modeling under the framework of link
prediction, which might be interested by network scientists
\cite{Lu2011,Wang2011}. In the practical aspect, for biological
networks such as protein-protein interaction networks and metabolic
networks \cite{Yu2008,Stumpf2008,Amaral2008}, the experiments of
uncovering the new links or interactions are costly, and thus to
predict in advance and focus on the links most likely to exist can
sharply reduce the experimental costs \cite{Clauset2008}. In
addition, some of the representative methods of link prediction have
been successfully applied to address the classification problem in
partially labeled networks \cite{Gallagher2008,Zhang2010}, as well
as identified the spurious links resulted from the inaccurate
information in the data \cite{Grumera2009}.

Motivated by the theoretical interests and practical significance,
many methods for link predication have been proposed. Therein some
algorithms are based on Markov
chains\cite{BilgicIEEE,Zhu2002,Sarukkai2000} and machine
learning\cite{Popescul2003,Wang2007}. Another group of algorithms
are based on node similarity \cite{Liben-Nowell2007,Lu2011}. Common
Neighbors (CN) is one of the simplest similarity indices.
Empirically, Kossinets and Watts\cite{Kossinets2006} analyzed a
large-scale social network, suggesting that two students having many
mutual friends are very probable to be friend in the future. Also in
online society like Facebook, the users tend to be friend if they
have common friends and therefore form a community. Extensive
analysis on disparate networks suggests that CN index is of good performance
on predicting the missing links \cite{Liben-Nowell2007,Zhou2009}.
L\"u \emph{et al.} \cite{LuPRE2009} suggested that in a network with
large clustering coefficient, CN can provide competitively accurate
predictions compared with the indices making use of global
information. Very recently, Cui \emph{et al.} \cite{Cui2011}
revealed that the nodes with more common neighbors are more likely
to form new links in a growing network. The basic assumption of CN
is that two nodes are more likely to be connected if they have more
common neighbors. Simply counting the number of common neighbors
indicates that each common neighbor gives equal contribution to the
connection likelihood. However, sometimes different common neighbors
may play different roles. For instance, the common close friends of
two people who don't know each other may contribute more to their
possibly future friendship than their common nodding acquaintances.
In this Letter, we propose a probabilistic model based on the
Bayesian theory, called Local Na\"ive Bayes (LNB) model, to predict
the missing links in complex networks. Based on the LNB model, two
node pairs with exactly the same number of common neighbors may have
much different connection likelihoods. Experiments on eight real
networks demonstrate that our method can effectively identify the
different roles of common neighbors to the connection likelihood and
thus give more accurate prediction than CN.

\section{Problem Description}
Consider an undirected network $G(V,E)$, where $V$ and $E$ are the sets of nodes and links, respectively. The multiple links and self-connections are not allowed. Each nonexistent link, namely a link $(x, y)\in U-E$ where $x, y\in V$ and $U$ denotes the universal set, will be assigned a score to quantify its existence likelihood. Higher score means higher probability that nodes $x$ and $y$ are connected, and vice versa. All the nonexistent links are sorted in descending order according to their scores, and the links at the top are most likely to exist. To test the algorithm's accuracy, the observed links, $E$, are randomly divided into two parts: the training set, $E^T$, is treated as known information, while the probe set, $E^P$, is used for testing and no information therein is allowed to be used for prediction. Clearly, $E = E^T\cup E^P$ and $E^T\cap E^P =\phi$. In this Letter, the training set always contains 90\% of links, and the remaining 10\% of links constitute the probe set. Hereinafter, the links in $E^P$ are called missing links and the links in $U-E^T$ are called non-observed links.

We apply two standard metrics to quantify the prediction accuracy: \emph{AUC} (area under the receiver operating characteristic curve) \cite{Hanley1982} and \emph{precision} \cite{Herlocker2004}. The AUC evaluates the algorithm¡¯s performance according to the whole list. Provided the rank of all non-observed links, AUC can be interpreted as the probability that a randomly chosen missing link is given a higher score than a randomly chosen nonexistent link. In the implementation, among $n$ times of independent comparisons, if there are $n'$ times the missing link having higher score and $n''$ times the missing link and nonexistent link having the same score, the AUC value is:
\begin{equation}
\texttt{AUC} = \frac{n'+0.5n''}{n}.
\end{equation}
If all the scores are generated from an independent and identical distribution, AUC should be about 0.5. Therefore, the degree to which the value exceeds 0.5 indicates how much better the algorithm performs than pure chance. Different from AUC, precision only focuses on the $L$ links with top ranks or highest scores. It is defined as the ratio of relevant items selected to the items selected. Among the top-$L$ links, if $L_r$ links are accurately predicted (i.e., there are $L_r$ links in the probe set), then the precision equals $L_r/L$. Clearly, higher precision means higher prediction accuracy.

\section{Method}
Given a pair of disconnected nodes ($x$,$y$) (i.e., a non-observed link), our task is to calculate the probability of connecting these two nodes on the basis of the condition: $x$ and $y$ might have a group of common neighbors, each of which has a couple of conditional probabilities corresponding to encouraging and hampering the connections of its two neighbors (see Eqs. (\ref{Eq_con_1}) and (\ref{Eq_con_0})).

\subsection{The Na\"ive Bayes Classifier}
A na\"ive Bayes classifier is a simple probabilistic classifier based on Bayesian theory with strong (na\"{i}ve) independence assumptions that the presence (or absence) of a particular feature of a class is unrelated to the presence (or absence) of any other feature \cite{naive-Bayes-Wiki}. Abstractly, the probability model for a classifier is a conditional model $P(C|F_1,\cdots,F_n)$ where $C$ is a dependent class variable and $F_1,F_2,\cdots,F_n$ are feature variables. According to the Bayesian theory\footnote{Bayesian theory \cite{bayesian,bayesWiki} is a probabilistic approach which relates the conditional and marginal probabilities of events $A$ and $B$, provided that the probability of $B$ does not equal to zero: $P(A|B)=\frac{P(A)\cdot P(B|A)}{P(B)}$, where $P(A)$ is the prior probability (or marginal probability) of $A$ which does not take into account any information about $B$, and $P(A|B)$ is the conditional probability of $A$ given $B$. $P(A|B)$ is also called the posterior probability.}, the posterior probability $P(C|F_1,F_2,\cdots,F_n)$ is
\begin{equation}
\label{Eq-bayes2}
P(C|F_1,F_2,\cdots,F_n)=\frac{P(C)\cdot P(F_1,F_2,\cdots,F_n|C)}{P(F_1,F_2,\cdots,F_n)}.
\end{equation}
Consider the na\"ive assumption that each feature $F_i$ is conditionally independent to every other feature $F_j$  ($j\neq i$), then we have
\begin{equation}
\label{Eq-bayes3}
P(C|F_1,F_2,\cdots,F_n)=\frac{P(C)\cdot \prod_{i=1}^{n}P(F_i|C)}{P(F_1,F_2,\cdots,F_n)}.
\end{equation}

\begin{table*}
\caption{The basic topological features of the eight networks. $|V|$
and $|E|$ are the number of nodes and links. $C$ and $r$ are
clustering coefficient\cite{Celegans} and assortative
coefficient\cite{Newman2002}, respectively. $\langle k\rangle$ is
the average degree of network. $\langle d\rangle$ is the average
shortest distance between node pairs. $H$ denotes the degree
heterogeneity defined as $H =\frac{\langle k^2\rangle}{\langle
k\rangle^2}$. \label{topology}} \centering
{\begin{tabular}{ccccccccc} \hline \hline
Networks &USAir &Yeast &CE &PB &NS  &FW1 &FW2 &FW3\\
\hline
$|V|$   &332 &2375 &297 &1222 &379 &128 &69 &97\\
$|E|$   &2126 &11693 &2148 &16717 &941 &2106 &880 &1446\\
$C$   &0.749 &0.388 &0.308 &0.361 &0.798 &0.335 &0.552 &0.468\\
$r$   &-0.208 &0.454 &-0.163 &-0.221 &-0.082 &-0.104 &-0.298 &-0.151\\
$\langle k\rangle$ &12.81 &9.85 &14.46 &27.36 &4.82 &32.90 &25.51 &29.81\\
$\langle d\rangle$ &2.46 &4.59 &2.46 &2.51 &4.93 &1.77 &1.64 &1.69\\
$H$ &3.46 &3.48 &1.80 &2.97 &1.663 &1.231 &2.33 &1.48\\
\hline \hline
\end{tabular}}
\end{table*}

\subsection{Local Na\"ive Bayes Model}
Given a training set $G(V,E^T)$, link prediction questions which links are more likely to exist among all the non-observed links. Denote by $A_1$ and $A_0$ the class variables of \emph{connection} and \emph{disconnection} respectively. The prior probabilities of $A_1$ and $A_0$ can be calculated through
\begin{equation}
\label{Eq_pri_1}
P(A_1)=\frac{M^T}{M},
\end{equation}
\begin{equation}
\label{Eq_pri_0}
P(A_0)=\frac{M-M^T}{M},
\end{equation}
where $M^T=|E^T|$ and $M=|U|=\frac{1}{2}|V|\cdot(|V|-1)$. Each node $w$ owns two conditional probabilities \{$P(w|A_1), P(w|A_0)$\}, where $P(w|A_1)$ is the probability that node $w$ is the common neighbor of two connected nodes, and $P(w|A_0)$ is the probability that node $w$ is the common neighbor of two disconnected nodes. According to Bayesian theory, these two probabilities are
\begin{equation}
\label{Eq_con_1}
P(w|A_1)=\frac{P(w)\cdot P(A_1|w)}{P(A_1)},
\end{equation}
\begin{equation}
\label{Eq_con_0}
P(w|A_0)=\frac{P(w)\cdot P(A_0|w)}{P(A_0)}.
\end{equation}
For a pair of disconnected nodes ($x$,$y$), denote by $O_{xy}$ the set of their common neighbors that are considered as the feature variables and assume they are independent to each other. Then according to the na\"ive Bayes classifier theory, the posterior probability of connection and disconnection of node $x$ and $y$ are
\begin{equation}
\label{Eq-cnbayes1}
P(A_1|O_{xy})=\frac{P(A_1)}{P(O_{xy})}\prod_{w\in O_{xy}}P(w|A_1),
\end{equation}
\begin{equation}
\label{Eq-cnbayes2}
P(A_0|O_{xy})=\frac{P(A_0)}{P(O_{xy})}\prod_{w\in O_{xy}}P(w|A_0).
\end{equation}
For a given node pair, comparing these two probabilities we can obtain whether they are likely to connect. However, it can not tell us which non-observed links are more likely to exist than the others. In order to compare the existence likelihood between the node pairs, we define the likelihood score of node pair ($x$,$y$) as the ratio of Eq. (\ref{Eq-cnbayes1}) to Eq. (\ref{Eq-cnbayes2}). Substituting Eqs. (\ref{Eq_con_1}) and (\ref{Eq_con_0}), we have
\begin{equation}
\label{Eq_Pre2}
r_{xy}=\frac{P(A_1)}{P(A_0)}\prod_{w\in O_{xy}}\frac{P(A_0)\cdot P(A_1|w)}{P(A_1)\cdot P(A_0|w)}.
\end{equation}
Indeed $P(A_1|w)$ is equal to the clustering coefficient of node $w$, denote by $C_w$ that can be calculated by
\begin{equation}
\label{PA1w}
P(A_1|w)=C_w=\frac{N_{\triangle w}}{N_{\triangle w}+N_{\wedge w}},
\end{equation}
where $N_{\triangle w}$ and $N_{\wedge w}$ are respectively the number of connected and disconnected node-pairs whose common neighbors include $w$. Obviously $N_{\triangle w}+N_{\wedge w}=\frac{k_w\times(k_w-1)}{2}$, where $k_w$ is the degree of node $w$. Since $P(A_1|w)+P(A_0|w)=1$, we have
\begin{equation}
\label{PA0w}
P(A_0|w)=1-C_w=\frac{N_{\wedge w}}{N_{\triangle w}+N_{\wedge w}}.
\end{equation}
Substituting Eqs. (\ref{Eq_pri_1}) (\ref{Eq_pri_0}) and Eqs. (\ref{PA1w}) (\ref{PA0w}), the likelihood score of node pair ($x$,$y$) is
\begin{equation}
\label{Eq_Pre3}
r_{xy}=s^{-1}\prod_{w\in O_{xy}}s\frac{N_{\triangle w}+1}{N_{\wedge w}+1}.
\end{equation}
where $s=\frac{P(A_0)}{P(A_1)}=\frac{M}{M^T}-1$ is a constant for a given training set, and thus $s^{-1}$ can be neglected in calculation. Note that we here apply the \emph{add-one smoothing} to prevent the score from being equal to $0$. Clearly, larger score means higher probability that the two nodes are connected. For a given node $w$, we directly define its role function as
\begin{eqnarray}
\label{eq-ratio}
R_w &=& \frac{N_{\triangle w}+1}{N_{\wedge w}+1}.
\end{eqnarray}
Therefore Eq. (\ref{Eq_Pre3}) can be written as
\begin{equation}
\label{rxy}
r_{xy}=\prod_{w\in O_{xy}}s R_w.
\end{equation}
Clearly, if $R_w=1$ for all nodes in the network, then the score of nodes $x$ and $y$, $r^{xy}$, will become a monotone increasing function of the number of their common neighbors. In this case, Eq. (\ref{rxy}) is equivalent to CN ($=|O_{xy}|$). Different common neighbors are generally of different contributions to the connecting probability, according to the corresponding $R_w$.

\begin{table*}
\caption {The prediction accuracy measured by AUC and precision (top-100) on eight networks. Each value is obtained
by averaging over 100 implementations with independently random
divisions of training set and probe set. \label{accuracy}}
\centering \rowcolors{2}{gray!20}{white}
\begin{tabular}{cccccccccc}
\hline \hline
       & Index   &USAir   &Yeast   &CE   &PB   &NS   &FW1 &FW2 &FW3\\
\hline
       \textbf{AUC}&CN      &0.953  &0.916 &0.848 &0.924 &0.980 &0.606 &0.689 &0.710\\
       &LNB-CN  &0.959  &0.916 &0.862 &0.926 &0.982 &0.694 &0.733 &0.748\\
       &AA      &0.965  &0.916 &0.865 &0.927 &0.984 &0.608 &0.697 &0.713\\
       &LNB-AA  &0.967  &0.916 &0.866 &0.928 &0.984 &0.697 &0.733 &0.750\\
       &RA      &0.972  &0.916 &0.870 &0.928 &0.984 &0.613 &0.704 &0.716\\
       &LNB-RA  &0.972  &0.917 &0.867 &0.929 &0.984 &0.697 &0.730 &0.750\\
\hline \hline
       \textbf{Precision}&CN  &0.597  &0.685 &0.131 &0.419 &0.356 &0.087 &0.146 &0.133\\
       &LNB-CN  &0.612  &0.689 &0.138 &0.409 &0.391 &0.106 &0.192 &0.161\\
       &AA   &0.615  &0.699 &0.135 &0.378 &0.527 &0.090 &0.156 &0.139\\
       &LNB-AA  &0.629  &0.703 &0.136 &0.380 &0.528 &0.104 &0.193 &0.161\\
       &RA      &0.630  &0.506 &0.126 &0.247 &0.547 &0.086 &0.169 &0.145\\
       &LNB-RA  &0.633  &0.625 &0.129 &0.259 &0.548 &0.104 &0.196 &0.170\\
\hline \hline
\end{tabular}
\end{table*}

It has been pointed out that the common neighbor's degree play an important role in link prediction. Suppressing the contributions of common neighbors with high degrees can improve the prediction accuracy \cite{Zhou2009}. Two indices are designed in this way, Adamic-Adar index (AA) \cite{Adamic2003} and Resource Allocation (RA) \cite{Zhou2009,Weiping2010}. The scores are defined respectively as:
\begin{equation}
r_{xy}^{\texttt{AA}}=\sum_{w\in O_{xy}}\frac{1}{\mathrm{log}k_w},
\end{equation}
and
\begin{equation}
r_{xy}^{\texttt{RA}}=\sum_{w\in O_{xy}}\frac{1}{k_w}.
\end{equation}
With the same motivation, we add an exponent $f(k_w)$ to the item $sR_w$ in Eq. (\ref{rxy}), where $f$ is a function of node's degree. Using Log function on both sides, we obtain a linear formula of connection likelihood:
\begin{equation}
r_{xy}'=\sum_{w\in O_{xy}}f(k_w)\mathrm{log}(sR_w).
\end{equation}
Here we consider three forms of function $f$, namely $f(k_w)=1$, $f(k_w)=\frac{1}{\mathrm{log}k_w}$ and $f(k_w)=\frac{1}{k_w}$, which are corresponding to the Local Na\"{i}ve Bayes (LNB) form of CN, AA and RA indices, respectively:
\begin{equation}
r_{xy}^\texttt{LNB-CN}=|O_{xy}|\mathrm{log}s+\sum_{w\in O_{xy}}\mathrm{log}R_w,
\end{equation}
\begin{equation}
r_{xy}^\texttt{LNB-AA}=\sum_{w\in O_{xy}}\frac{1}{\mathrm{log}k_w}(\mathrm{log}s+\mathrm{log}R_w),
\end{equation}
\begin{equation}
r_{xy}^\texttt{LNB-RA}=\sum_{w\in O_{xy}}\frac{1}{k_w}(\mathrm{log}s+\mathrm{log}R_w).
\end{equation}
Obviously, when $R_w=1$, namely we don't consider the differen roles of common neighbors, the LNB-CN, LNB-AA and LNB-RA will degenerate to CN, AA and RA, respectively.

\section{Results}
Eight networks are considered in our experiments: (i) USAir \cite{USAir}: The network of US air transportation system, which contains 332 airports and 2126 airlines. (ii) C.elegans (CE)  \cite{Celegans}: The neural network of the nematode worm C.elegans, in which an edge joins two neurons if they are connected by either a synapse or a gap junction. This network contains 297 neurons and 2148 links. (iii) Political Blogs (PB)  \cite{PB}: A network of the US political blogs. The original links are directed, here we treat them as undirected links. (iv) Yeast  \cite{Yeast}: A protein-protein interaction network containing 2617 proteins and 11855 interactions. Although this network is not well connected (it contains 92 components), most of nodes belong to the giant component, whose size is 2375. (v) NetScience (NS)  \cite{Newman2006}: A network of coauthorships between scientists who are themselves publishing on the topic of networks. The network contains 1589 scientists, and 128 of which are isolated. Here we do not consider those isolated nodes. The connectivity of NS is not good, actually, NS is consisted of 268 connected components, and the size of the largest connected component is only 379. (vi) Foodweb1 (FW1) \cite{FW1}: A network of foodweb in Florida Bay during wet season. (vii) Foodweb2 (FW2) \cite{FW2}: A network of foodweb in Everglades Graminoids during wet season. (viii) Foodweb3 (FW3) \cite{FW3}: A network of foodweb in Mangrove Estuary during wet season. Here we only consider the giant component. The basic topological features of these eight networks are summarized in Table \ref{topology}.

The prediction accuracy, measured by AUC and precision, on the eight real networks are shown in Table \ref{accuracy}. In general, the LNB forms outperform their corresponding basic forms. It shows that for AUC, except the result of RA in CE network, LNB model gives higher accurate prediction for all eight networks. Especially the improvements of the foodwebs are significant. When measured by precision, except the result of CN in PB network, LNB model also improve the accuracy. In accordance with our analysis, CN method assigns equal weight to the common neighbors (i.e., $R_w=1$ for all nodes), while LNB-CN can effectively capture the different roles of common neighbors.

\section{Case Study}
In this section, we give detailed analysis on USAir network which contains 332 airports and 2126 airlines. This network has a very specific structure: the hierarchical organization consisting of hubs, local centers and small local airports. We rank all the airports according to their degrees in descending order. The top 17 airports who own about 1/3 of the total degree are defined as hubs (Hub), the last 273 airports who also own 1/3 of the total degree are local airports (LA), and the rest 42 airports are local centers (LC). Therefore there are six kinds of links: Hub-Hub (97.06\%), Hub-LC (72.83\%), LC-LC (36.24\%), LA-Hub (12.5\%), LA-LC (2.75\%) and LA-LA (0.72\%). The number in the bracket indicates the connecting probability: the ratio of the number of real links to its possibly maximal value. For example, there are 132 links connecting two hubs among $\frac{17\times16}{2}=136$ possible links, and thus the connection probability for two hubs is 97.06\%, indicating that hubs are densely connected with each other, which is a specific feature of air transportation network \cite{Colizza2006}. Due to this reason, both CN and LNB-CN methods can provide accurate predictions on the missing Hub-Hub links. Moveover, we find that although the number of common neighbors of (LC,LC) are higher than that of (Hub,LC), LCs are more likely to connect with Hubs (see $72.83\%>26.24\%$). Therefore by simply counting the number of common neighbors, CN tends to assign higher score to (LC,LC) than (Hub,LC) and thus leads to poor predictions. Since LBN-CN is sophisticate to identify the negative roles of (LC,LC)'s common neighbors, it will depress the score of (LC,LC) and provide more accurate predictions on Hub-LC and LC-LC links. Compared with the first three kinds of links, the rest three kinds of links involving local airports are difficult to predict (see later that none of such links are include in top-100 with both CN and LNB-CN). Because LAs usually connected with Hubs and rarely have common neighbors with other airports, the nodes pairs involving LAs are given very small scores and thus ranked lower. This is a common drawback of CN-based methods.

\begin{figure}
\begin{center}
\includegraphics[width=6.8cm]{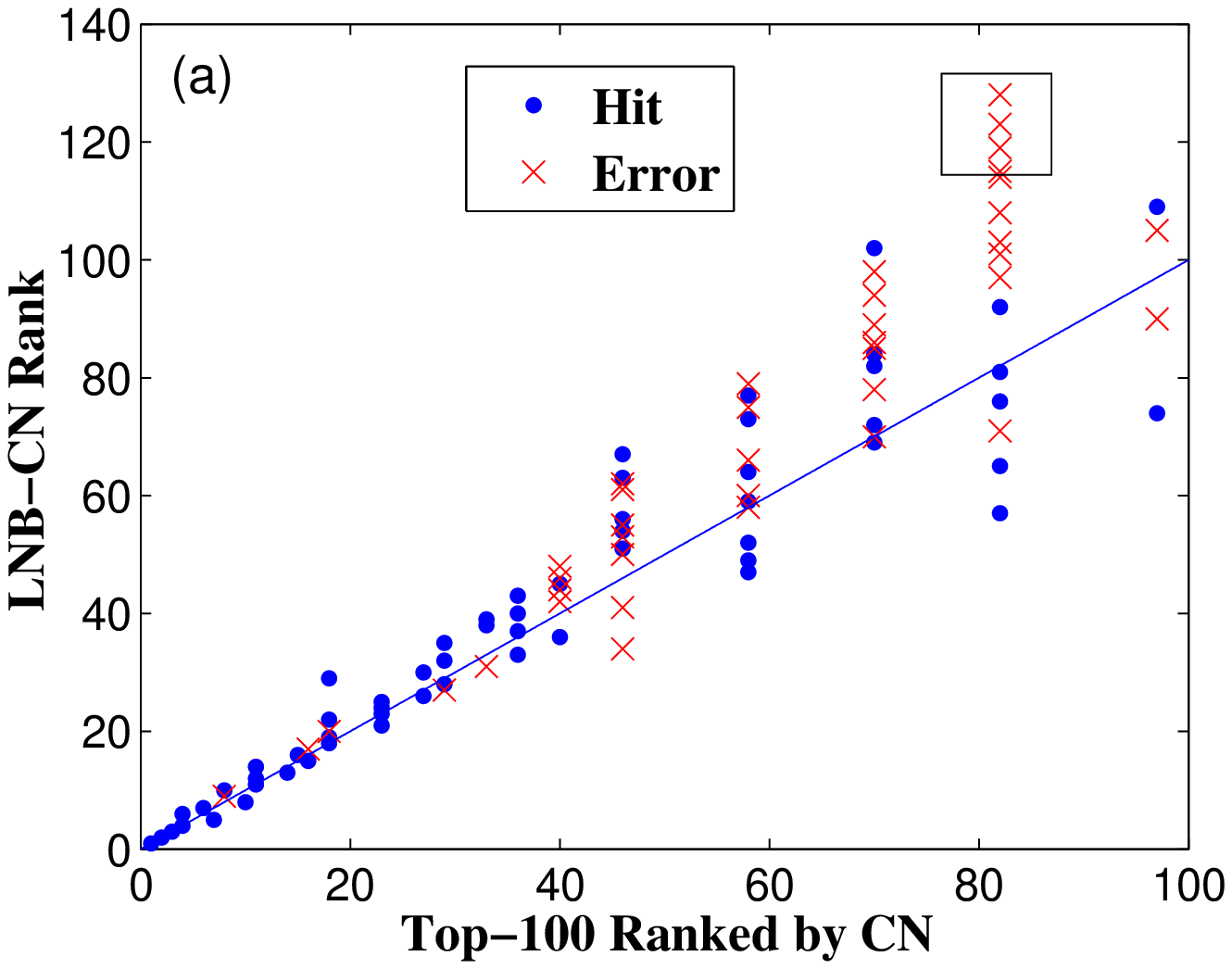}
\includegraphics[width=6.8cm]{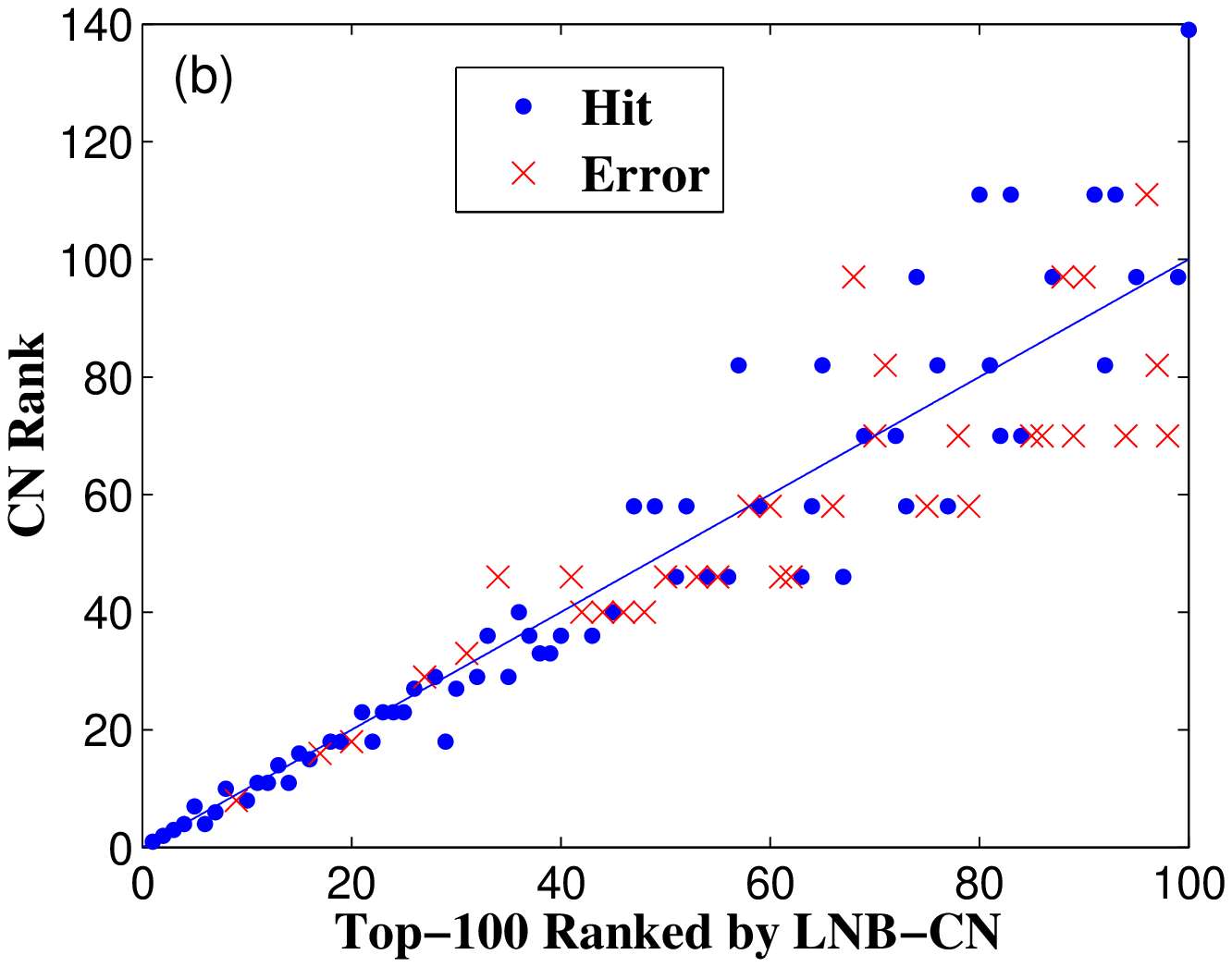}
\caption{Comparison of the top-100 node pairs respectively ranked by CN and LNB-CN on USAir network. (a) The top-100 node pairs ranked by CN and their corresponding ranks assigned by LNB-CN. (b) The top-100 node pairs ranked by LNB-CN and their corresponding ranks assigned by CN. ``Hit" denotes the accurate predicted link (i.e., links in probe set), while ``Error" indicates the non-existent link. The diagonal is presented by a blue solid line.\label{Toplist}}
\end{center}
\end{figure}

We further focus on the top-100 node pairs respectively ranked by CN and LNB-CN. Fig. \ref{Toplist}(a) shows the top-100 node pairs ranked by CN and their corresponding ranks assigned by LNB-CN, and Fig. \ref{Toplist}(b) shows the top-100 node pairs ranked by LNB-CN and their corresponding ranks assigned by CN. The links who are accurately predicted (i.e., $\in E^P$) are labeled by blue dots while the non-existent links (i.e., $\in U-E$) are labeled by red crosses. From Fig. \ref{Toplist}(a), we can see that among top-100 node pairs ranked by CN, 11 node pairs are ranked above 100 by LNB-CN, within which only 2 of them are predicted right, which implies that LNB-CN tends to rank the non-existent links lower than CN. Among the rest 9 node pairs, 6 of them are LC-LC links and 3 pairs are of type Hub-LC. This result further demonstrates that LNB-CN can give more accurate judgements on the LC-LC and Hub-LC links. Similarly, in Fig. \ref{Toplist}(b), 6 pairs are ranked above 100 by CN, within which 5 links are predicted right, which indicates that LNB-CN tends to rank missing links higher than CN.

\begin{figure}
\begin{center}
\includegraphics[width=4.3cm]{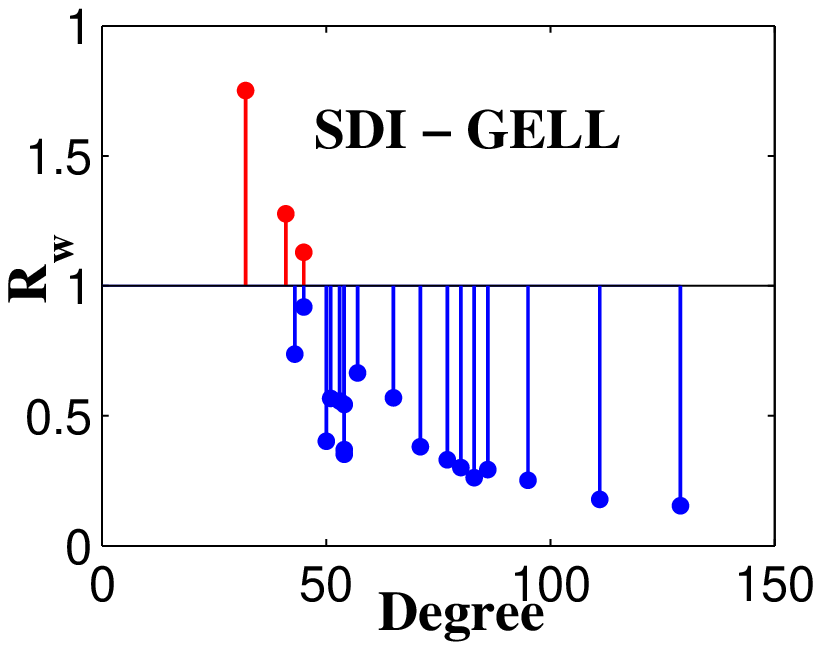}
\includegraphics[width=4.3cm]{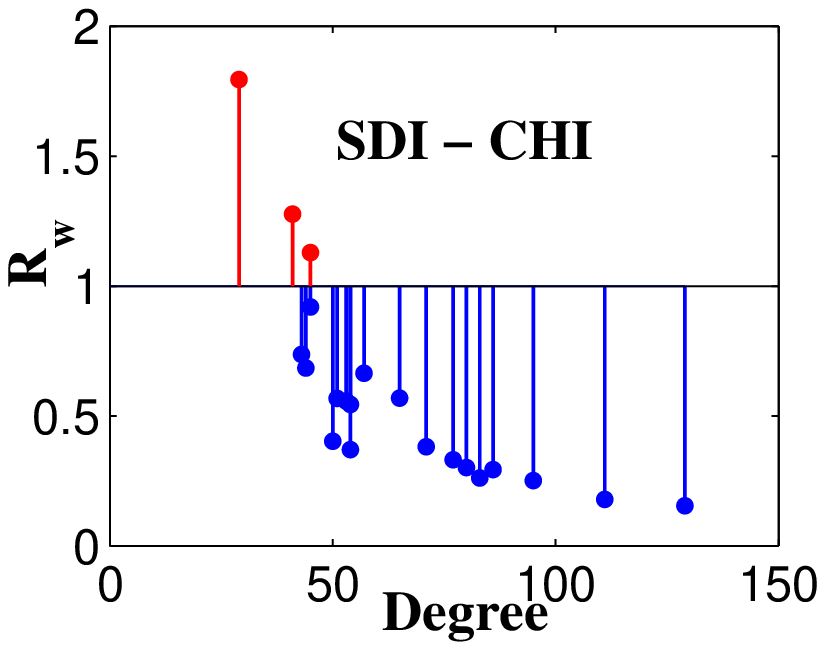}
\includegraphics[width=4.3cm]{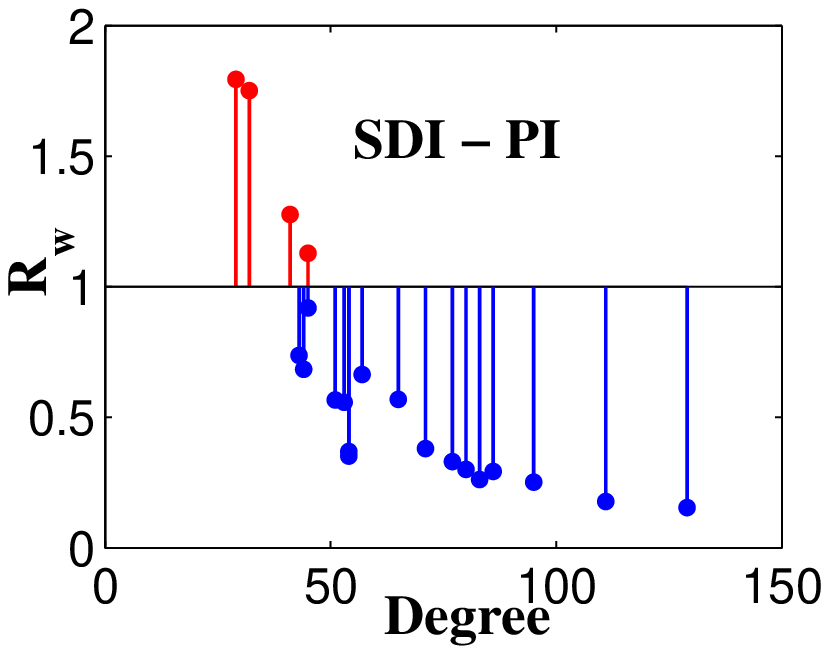}
\includegraphics[width=4.3cm]{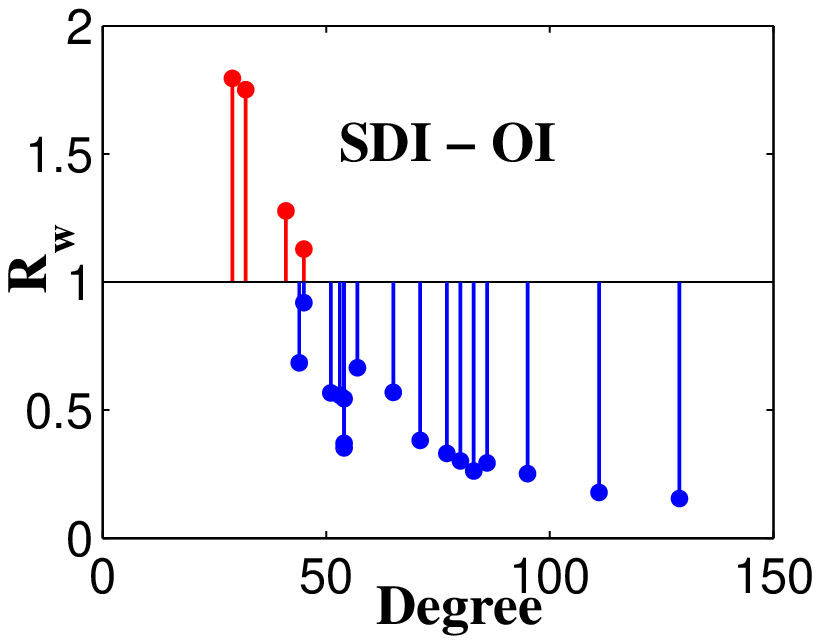}
\caption{The dependence of common neighbors' $R_w$ on its degree for the four pairs.\label{cnpr}}
\end{center}
\end{figure}

We take four pairs within the rectangle in Fig. \ref{Toplist} (a) as examples. Five airports are involved, including \emph{San Diego Intl} (SDI), \emph{General Edward Lawrence Logan} (GELL), \emph{Cleveland-Hopkins Intl} (CHI), \emph{Philadelphia Intl} (PI) and \emph{Orlando Intl} (OI). Therein, PI, with over 62 airlines is a hub, others are local airports. These four pairs, ranked 128, 123, 119 and 114 respectively by LNB-CN, are corresponding to four non-existent airlines, namely (GELL, SDI), (CHI, SDI), (OI, SDI) and (PI, SDI). Each of them has 21 common neighbors and thus be ranked as top-82 by CN. Since SDI is located at the west coast while other three airports are all at the east of USA, there are no direct airlines connecting SDI and other three airports. Instead the passengers need to transfer at their common neighbors (most of which are hubs). This feature can be well captured by LNB-CN. Figure \ref{cnpr} shows the dependence of common neighbors' $R_w$ on its degree for these four pairs. Clearly, for all four pairs most of the common neighbors play negative roles and thus hamper the connections. Therefore, LNB-CN will rank this kind of pairs lower than CN by assigning them small scores.

\section{Conclusion}

In this Letter, we proposed a local naive Bayes model (LNB) to
predict missing links in complex networks. The advantage of this
method is that it can well capture the different roles of common
neighbors and assign them different weights. To test the method, we
compared three representative local indices, namely Common Neighbors
(CN), Adamic-Adar (AA) index and Resource Allocation (RA) index,
with their corresponding LNB forms. Extensive analysis on eight real
networks drown from disparate fields shows that for all these three
indices the LNB forms outperform their corresponding original
indices. In particular, the improvement is remarkable on the
foodwebs where the hierarchical structure is obvious and there are
rare links within the same level. Finally, we gave a detailed
analysis on US air transportation network. Although some pairs of
airports have many common neighbors, there are no directed airlines
connecting them because of the long geographical distance. LNB
methods can well capture this feature and thus give more accurate
predictions. In addition, some researchers found that the local
clustering property can be utilized to improve the accuracy of link
prediction~\cite{Feng2011}, yet they did not give any solid reason
about their method, while the present model provides a theoretical
base on the usage of local clustering (see Eq. (14)).

\acknowledgments
We acknowledge Chong-Jing Sun for helpful assistance in manuscript preparation and Duanbing Chen for valuable suggestions. This work is partially supported by the National Natural Science Foundation of China under Grant Nos. 11075031, 60903073 and 60973069. L.L. acknowledges the Swiss National Science Foundation under Grant No. 200020-132253.

\end{document}